\documentclass[letterpaper,english,conference, twocolumn]{IEEEtran}
\usepackage[T1]{fontenc}
\usepackage[latin9]{inputenc}
\usepackage{verbatim}
\usepackage{float}
\usepackage{amsthm}
\usepackage{amsmath}
\usepackage{amssymb}
\usepackage{esint}
\PassOptionsToPackage{normalem}{ulem}
\usepackage{ulem}

\makeatletter

\floatstyle{ruled}
\newfloat{algorithm}{tbp}{loa}
\providecommand{\algorithmname}{Algorithm}
\floatname{algorithm}{\protect\algorithmname}

\theoremstyle{definition}
\newtheorem{assumption}{Assumption}
  \theoremstyle{plain}
  \newtheorem{thm}{\protect\theoremname}
  \theoremstyle{remark}
  \newtheorem{rem}{\protect\remarkname}

\allowdisplaybreaks
\usepackage{cite}
\IEEEoverridecommandlockouts
\makeatother

\usepackage{babel}
\providecommand{\remarkname}{Remark}
\providecommand{\theoremname}{Theorem}

\begin{document}

\title{Concurrent learning-based online approximate feedback-Nash equilibrium
solution of $N$-player nonzero-sum differential games%
\thanks{Rushikesh Kamalapurkar, Justin Klotz, and Warren E. Dixon are with
the Department of Mechanical and Aerospace Engineering, University
of Florida, Gainesville, FL, USA. Email: \{rkamalapurkar, jklotz,
wdixon\}@ufl{}.edu.%
}%
\thanks{This research is supported in part by NSF award numbers 0901491, 1161260,
1217908, ONR grant number N00014-13-1-0151, and a contract with the
AFRL Mathematical Modeling and Optimization Institute. Any opinions,
findings and conclusions or recommendations expressed in this material
are those of the authors and do not necessarily reflect the views
of the sponsoring agency.%
}}

\author{Rushikesh Kamalapurkar, Justin Klotz, and Warren E. Dixon}
\maketitle
\begin{abstract}
This paper presents a concurrent learning-based actor-critic-identifier
architecture to obtain an approximate feedback-Nash equilibrium solution
to an infinite horizon $N$-player nonzero-sum differential game online,
without requiring persistence of excitation (PE), for a nonlinear
control-affine system. Under a condition milder than PE, uniformly
ultimately bounded convergence of the developed control policies to
the feedback-Nash equilibrium policies is established.
\end{abstract}

\section{Introduction}

Various control problems can be modeled as multi-input systems, where
each input is computed by a player, and each player attempts to influence
the system state to minimize its own cost function. In this case,
the optimization problem for each player is coupled with the optimization
problem for other players. In general, an optimal solution in the
usual sense does not exist; and hence, alternative criteria for optimality
are sought.

Differential game theory provides solution concepts for multi-player,
multi-objective optimization problems \cite{Isaacs1999,Tijs2003,Basar1999}.
For example, a set of policies is called a Nash equilibrium solution
to a multi-objective optimization problem if none of the players can
improve their outcome by changing their policy if all the other players
abide by the Nash equilibrium policies \cite{Nash1951}. The Nash
equilibrium provides a secure set of strategies, in the sense that
none of the players have an incentive to diverge from their equilibrium
policy. Hence, Nash equilibrium has been a widely used solution concept
in differential game-based control techniques. 

In general, Nash equilibria are not unique. For a closed-loop differential
game (i.e., the control is a function of the state and time) with
perfect information (i.e. all the players know the complete state
history), there can be infinitely many Nash equilibria. However, if
the policies are constrained to be feedback policies, the resulting
equilibria are called (sub)game-perfect-Nash equilibria or feedback-Nash
equilibria. The value functions corresponding to feedback-Nash equilibria
satisfy a coupled system of Hamilton-Jacobi (HJ) equations \cite{Case1969,Starr.Ho1969,Starr1969,Friedman1971}. 

If the system dynamics are nonlinear and uncertain, an analytical
solution of the coupled HJ equations is generally infeasible. Hence,
dynamic programming-based approximate solutions are sought \cite{Littman2001,Wei2008,Zhang2010,Zhang2010a,Vrabie2010,Vamvoudakis2011}.
In \cite{Vrabie2010}, an integral reinforcement learning algorithm
is presented to solve nonzero-sum differential games in linear systems
without the knowledge of the drift matrix. In \cite{Vamvoudakis2011},
a dynamic programming-based technique is developed to find an approximate
feedback-Nash equilibrium solution to an infinite horizon $N$-player
nonzero-sum differential game online for nonlinear control-affine
systems with known dynamics. In \cite{Johnson2011a}, a policy iteration-based
method is used to solve a two-player zero-sum game online for nonlinear
control-affine systems without the knowledge of drift dynamics.

The methods in \cite{Vamvoudakis2011} and \cite{Johnson2011a} solve
the differential game online using a parametric function approximator
such as a neural network (NN) to approximate the value functions.
Since the approximate value functions do not satisfy the coupled HJ
equations, a set of residual errors (the so-called Bellman errors
(BEs)) is computed along the state trajectories and is used to update
the estimates of the unknown parameters in the function approximator
using least-squares or gradient-based techniques. Similar to adaptive
control, a restrictive persistence of excitation (PE) condition is
then used to ensure boundedness and convergence of the value function
weights. An ad-hoc exploration signal is added to the control signal
during the learning phase to satisfy the PE condition along the system
trajectories \cite{Mehta.Meyn2009,Vrabie2007,Sutton1998}.

Based on the ideas in recent concurrent learning-based results in
adaptive control such as \cite{Chowdhary.Yucelen.ea2012} and \cite{Chowdhary.Johnson2011a}
which show that a concurrent learning-based adaptive update law can
exploit recorded data to augment the adaptive update laws to establish
parameter convergence under conditions milder than PE, this paper
extends the work in \cite{Vamvoudakis2011} and \cite{Johnson2011a}
to relax the PE condition. In this paper, a concurrent learning-based
actor-critic architecture (cf. \cite{Bhasin.Kamalapurkar.ea2013a})
is used to obtain an approximate feedback-Nash equilibrium solution
to an infinite horizon $N$-player nonzero-sum differential game online,
without requiring PE, for a nonlinear control-affine system. 

The solutions to the coupled HJ equations and the corresponding feedback-Nash
equilibrium policies are approximated using parametric universal function
approximators. Using the known system dynamics, the Bellman errors
are evaluated at a set of preselected points in the state-space. The
value function and the policy weights are updated using a concurrent
learning-based least squares approach to minimize the instantaneous
BEs and the BEs evaluated at preselected points. It is shown that
under a condition milder than PE, uniformly ultimately bounded (UUB)
convergence of the value function weights and the policy weights to
their true values can be established.

\section{Problem formulation and exact solution}

Consider a class of control-affine multi-input systems
\begin{equation}
\dot{x}=f\left(x\right)+\sum_{i=1}^{N}g_{i}\left(x\right)\hat{u}_{i},\label{eq:Dynamics}
\end{equation}
where $x\in\mathbb{R}^{n}$ is the state and $\hat{u}_{i}\in\mathbb{R}^{m_{i}}$
are the control inputs (i.e. the players). In (\ref{eq:Dynamics}),
the functions $g_{i}:\mathbb{R}^{n}\to\mathbb{R}^{n\times m_{i}}$
are known, uniformly bounded, and locally Lipschitz, the function
$f:\mathbb{R}^{n}\to\mathbb{R}^{n}$ is known and $f\left(0\right)=0$.
Let $U\triangleq\{\left\{ u_{i}:\mathbb{R}^{n}\to\mathbb{R}^{m_{i}},i=1,..,N\right\} \mid\mbox{The tuple }\left\{ u_{i},..,u_{N}\right\} \mbox{ is admissible w.r.t. \eqref{eq:Dynamics}}\}$
be the set of admissible tuples of feedback policies. Let $V_{i}^{\left\{ u_{i},..,u_{N}\right\} }:\mathbb{R}^{n}\to\mathbb{R}_{\geq0}$
denote the value function of the $i^{th}$ player w.r.t. the tuple
of feedback policies $\left\{ u_{1},..,u_{N}\right\} \in U$, defined
as
\begin{equation}
V_{i}^{\left\{ u_{1},..,u_{N}\right\} }\left(x_{o}\right)=\intop_{t_{o}}^{\infty}r_{i}\left(x\left(\tau\right),u_{i}\left(x\left(\tau\right)\right),..,u_{N}\left(x\left(\tau\right)\right)\right)d\tau,\label{eq:Vi}
\end{equation}
where $x\left(\tau\right)$ for $\tau\in\mathbb{R}_{\geq0}$ denotes
the trajectory of (\ref{eq:Dynamics}) obtained using the feedback
policies $\hat{u}_{i}\left(\tau\right)=u_{i}\left(x\left(\tau\right)\right)$
and the initial condition $x\left(t_{o}\right)=x_{o}$. In (\ref{eq:Vi}),
$r_{i}:\mathbb{R}^{n}\times\mathbb{R}^{m_{1}}\times\cdots\times\mathbb{R}^{m_{N}}\to\mathbb{R}_{\geq0}$
denote the instantaneous costs defined as $r_{i}\left(x,u_{i},..,u_{N}\right)\triangleq x^{T}Q_{i}x+\sum_{j=1}^{N}u_{j}^{T}R_{ij}u_{j}$,
where $Q_{i}\in\mathbb{R}^{n\times n}$ are positive definite matrices.
The control objective is to find an approximate feedback-Nash equilibrium
solution to the infinite horizon regulation differential game online,
i.e., to find a tuple $\left\{ u_{1}^{*},..,u_{N}^{*}\right\} \in U$
such that for all $i\in\left\{ 1,..,N\right\} $, for all $x_{o}\in\mbox{\ensuremath{\mathbb{R}}}^{n}$,
the corresponding value functions satisfy
\[
V_{i}^{*}\left(x_{o}\right)\triangleq V_{i}^{\left\{ u_{1}^{*},u_{2}^{*},..,u_{i}^{*},..,u_{N}^{*}\right\} }\left(x_{o}\right)\leq V_{i}^{\left\{ u_{1}^{*},u_{2}^{*},..,u_{i},..,u_{N}^{*}\right\} }\left(x_{o}\right)
\]
 for all $u_{i}$ such that $\left\{ u_{1}^{*},u_{2}^{*},..,u_{i},..,u_{N}^{*}\right\} \in U$. 

The exact closed-loop feedback-Nash equilibrium solution $\left\{ u_{i}^{*},..,u_{N}^{*}\right\} $
can be expressed in terms of the value functions as \cite{Vamvoudakis2011,Basar1999,Starr.Ho1969,Starr1969}
\begin{equation}
u_{i}^{*}=-\frac{1}{2}R_{ii}^{-1}g_{i}^{T}\left(\nabla_{x}V_{i}^{*}\right)^{T},\label{eq:ui*}
\end{equation}
assuming that the solutions $\left\{ V_{1}^{*},..,V_{N}^{*}\right\} $
to the coupled Hamilton-Jacobi (HJ) equations 
\begin{multline}
x^{T}Q_{i}x+\sum_{j=1}^{N}\frac{1}{4}\nabla_{x}V_{j}^{*}G_{ij}\left(\nabla_{x}V_{j}^{*}\right)^{T}+\nabla_{x}V_{i}^{*}f\\
-\frac{1}{2}\nabla_{x}V_{i}^{*}\sum_{j=1}^{N}G_{j}\left(\nabla_{x}V_{j}^{*}\right)^{T}=0\label{eq:HJBCL*}
\end{multline}
exist and are continuously differentiable. In (\ref{eq:HJBCL*}),
$G_{j}\triangleq g_{j}R_{jj}^{-1}g_{j}^{T}$ and $G_{ij}\triangleq g_{j}R_{jj}^{-1}R_{ij}R_{jj}^{-1}g_{j}^{T}$.
The HJ equations in (\ref{eq:HJBCL*}) are in the so-called closed-loop
form; they can also be expressed in an open-loop form as
\begin{equation}
x^{T}Q_{i}x+\sum_{j=1}^{N}u_{j}^{*T}R_{ij}u_{j}^{*}+\nabla_{x}V_{i}^{*}f+\nabla_{x}V_{i}^{*}\sum_{j=1}^{N}g_{j}u_{j}^{*}=0.\label{eq:HJBOL*}
\end{equation}

\section{Approximate solution}

Computation of an analytical solution to the coupled nonlinear HJ
equations in (\ref{eq:HJBCL*}) is, in general, infeasible. Hence,
an approximate solution $\left\{ \hat{V}_{1},..,\hat{V}_{N}\right\} $
is sought. Based on $\left\{ \hat{V}_{1},..,\hat{V}_{N}\right\} $,
an approximation $\left\{ \hat{u}_{i},..,\hat{u}_{N}\right\} $ to
the closed-loop feedback-Nash equilibrium solution is determined.
Since the approximate solution, in general, does not satisfy the HJ
equations, a set of residual errors (the so-called Bellman errors
(BEs)) is computed as 
\begin{equation}
\delta_{i}=x^{T}Q_{i}x+\sum_{j=1}^{N}\hat{u}_{j}^{T}R_{ij}\hat{u}_{j}+\nabla_{x}\hat{V}_{i}f+\nabla_{x}\hat{V}_{i}\sum_{j=1}^{N}g_{j}\hat{u}_{j},\label{eq:BE1}
\end{equation}
and the approximate solution is recursively improved to drive the
BEs to zero.

\subsection{Value function approximation\label{sub:Value-function-approximation}}

Using the universal approximation property of NNs, the value functions
can be represented as
\begin{equation}
V_{i}^{*}\left(x\right)=W_{i}^{T}\sigma_{i}\left(x\right)+\epsilon_{i}\left(x\right),\label{eq:Vi*NN}
\end{equation}
where $W_{i}\in\mathbb{R}^{p_{W_{i}}}$ denote constant vectors of
unknown NN weights, $\sigma_{i}:\mathbb{R}^{n}\to\mathbb{R}^{p_{W_{i}}}$
denote the known NN activation functions, $p_{Wi}\in\mathbb{N}$ denote
the number of hidden layer neurons, and $\epsilon_{i}:\mathbb{R}^{n}\to\mathbb{R}$
denote the unknown function reconstruction errors. The universal function
approximation property guarantees that over any compact domain $\mathcal{C}\subset\mathbb{R}^{n}$,
for all constant $\overline{\epsilon}_{i},\overline{\epsilon}_{i}^{\prime}>0$,
there exists a set of weights and basis functions such that $\left\Vert W_{i}\right\Vert \leq\overline{W}$,
$\sup_{x\in\mathcal{C}}\left\Vert \sigma_{i}\left(x\right)\right\Vert \leq\overline{\sigma}_{i}$,
$\sup_{x\in\mathcal{C}}\left\Vert \sigma_{i}^{\prime}\left(x\right)\right\Vert \leq\overline{\sigma}_{i}^{\prime}$,
$\sup_{x\in\mathcal{C}}\left\Vert \epsilon_{i}\left(x\right)\right\Vert \leq\overline{\epsilon}_{i}$
and $\sup_{x\in\mathcal{C}}\left\Vert \epsilon_{i}^{\prime}\left(x\right)\right\Vert \leq\overline{\epsilon}_{i}^{\prime}$,
where $\overline{W}_{i},\overline{\sigma}_{i},\overline{\sigma}_{i}^{\prime},\overline{\epsilon}_{i},\overline{\epsilon}_{i}^{\prime}\in\mathbb{R}$
are positive constants. Based on (\ref{eq:ui*}) and (\ref{eq:Vi*NN}),
the feedback-Nash equilibrium solutions are
\begin{equation}
u_{i}^{*}\left(x\right)=-\frac{1}{2}R_{ii}^{-1}g_{i}^{T}\left(x\right)\left(\sigma_{i}^{\prime T}\left(x\right)W_{i}+\epsilon_{i}^{\prime T}\left(x\right)\right).\label{eq:ui*NN}
\end{equation}

The NN-based approximations to the value functions and the controllers
are defined as
\begin{gather}
\hat{V}_{i}\triangleq\hat{W}_{ci}^{T}\sigma_{i},\quad\hat{u}_{i}\triangleq-\frac{1}{2}R_{ii}^{-1}g_{i}^{T}\sigma_{i}^{\prime T}\hat{W}_{ai},\label{eq:Vu}
\end{gather}
where $\hat{W}_{ci}\in\mathbb{R}^{p_{W_{i}}}$, i.e., the value function
weights, and $\hat{W}_{ai}\in\mathbb{R}^{p_{W_{i}}}$, i.e., the policy
weights, are the estimates of the ideal weights $W_{i}$. The use
of two different sets of estimates to approximate the same set of
ideal weights is motivated by the subsequent stability analysis and
the fact that it facilitates an approximation of the BEs that is affine
in the value function weights, enabling least squares-based adaptation.
Based on (\ref{eq:Vu}), measurable approximations to the BEs in (\ref{eq:BE1})
are developed as
\begin{multline}
\hat{\delta}_{i}=\omega_{i}^{T}\hat{W}_{ci}+x^{T}Q_{i}x+\sum_{j=1}^{N}\frac{1}{4}\hat{W}_{aj}^{T}\sigma_{j}^{\prime}G_{ij}\sigma_{j}^{\prime T}\hat{W}_{aj},\label{eq:deltaiHat}
\end{multline}
where $\omega_{i}\triangleq\sigma_{i}^{\prime}f-\frac{1}{2}\sum_{j=1}^{N}\sigma_{i}^{\prime}G_{j}\sigma_{j}^{\prime T}\hat{W}_{aj}$.
The following assumption, which in general is weaker than the PE assumption,
is required for convergence of the concurrent learning-based value
function weight estimates.%

\begin{assumption}
\label{ass:CLW}For each $i\in\left\{ 1,..,N\right\} $, there exists
a finite set of $M_{xi}$ points $\left\{ x_{ij}\in\mathbb{R}^{n}\mid j=1,..,M_{xi}\right\} $
such that for all $t\in\mathbb{R}_{\geq0}$, 
\begin{align}
\mbox{rank}\left(\sum_{k=1}^{M_{ix}}\frac{\omega_{i}^{k}\left(t\right)\left(\omega_{i}^{k}\right)^{T}\left(t\right)}{\rho_{i}^{k}\left(t\right)}\right) & =p_{W_{i}},\nonumber \\
\underline{c}_{xi}\triangleq\frac{\left(\inf_{t\in\mathbb{R}_{\geq0}}\left(\lambda_{\min}\left\{ \sum_{k=1}^{M_{xi}}\frac{\omega_{i}^{k}\left(t\right)\omega_{i}^{kT}\left(t\right)}{\rho_{i}^{k}\left(t\right)}\right\} \right)\right)}{M_{xi}} & >0,\label{eq:CLRank2}
\end{align}
where $\lambda_{min}$ denotes the minimum eigenvalue, and $\underline{c}_{xi}\in\mathbb{R}$
are positive constants. In (\ref{eq:CLRank2}), $\omega_{i}^{k}\left(t\right)\triangleq\sigma_{i}^{\prime ik}f^{ik}-\frac{1}{2}\sum_{j=1}^{N}\sigma_{i}^{\prime ik}G_{j}^{ik}\left(\sigma_{j}^{\prime ik}\right)^{T}\hat{W}_{aj}\left(t\right)$
and $\rho_{i}^{k}\triangleq1+\nu_{i}\left(\omega_{i}^{k}\right)^{T}\Gamma_{i}\omega_{i}^{k}$,
where the superscripts $ik$ indicate that the terms are evaluated
at $x=x_{ik}$.
\end{assumption}
The concurrent learning-based least-squares update laws for the value
function weights are designed as
\begin{gather}
\dot{\hat{W}}_{ci}=-\eta_{c1i}\Gamma_{i}\frac{\omega_{i}}{\rho_{i}}\hat{\delta}_{i}-\frac{\eta_{c2i}\Gamma_{i}}{M_{xi}}\sum_{k=1}^{M_{xi}}\frac{\omega_{i}^{k}}{\rho_{i}^{k}}\hat{\delta}_{i}^{k},\nonumber \\
\dot{\Gamma}_{i}=\left(\beta_{i}\Gamma_{i}-\eta_{c1i}\Gamma_{i}\frac{\omega_{i}\omega_{i}^{T}}{\rho_{i}^{2}}\Gamma_{i}\right)\mathbf{1}_{\left\{ \left\Vert \Gamma_{i}\right\Vert \leq\overline{\Gamma}_{i}\right\} },\:\left\Vert \Gamma_{i}\left(t_{0}\right)\right\Vert \leq\overline{\Gamma}_{i},\label{eq:CriticUpdate}
\end{gather}
where $\rho_{i}\triangleq1+\nu_{i}\omega_{i}^{T}\Gamma_{i}\omega_{i}$,
$\mathbf{1}_{\left\{ \cdot\right\} }$ denotes the indicator function,
$\overline{\Gamma}_{i}>0\in\mathbb{R}$ are the saturation constants,
$\beta_{i}\in\mathbb{R}$ are the constant positive forgetting factors,
$\eta_{c1i},\eta_{c2i}\in\mathbb{R}$ are constant positive adaptation
gains, and the approximate BEs $\hat{\delta}_{i}^{k}$ are defined
as $\hat{\delta}_{i}^{k}\triangleq\left(\omega_{i}^{k}\right)^{T}\hat{W}_{ci}+x_{ik}^{T}Q_{i}x_{ik}+\sum_{j=1}^{N}\frac{1}{4}\hat{W}_{aj}^{T}\sigma_{j}^{\prime ik}G_{ij}^{ik}\left(\sigma_{j}^{\prime ik}\right)^{T}\hat{W}_{aj}.$

The policy weight update laws are designed based on the subsequent
stability analysis as
\begin{multline}
\dot{\hat{W}}_{ai}=-\eta_{a1i}\left(\hat{W}_{ai}-\hat{W}_{ci}\right)-\eta_{a2i}\hat{W}_{ai}\\
+\frac{1}{4}\sum_{j=1}^{N}\eta_{c1i}\sigma_{j}^{\prime}G_{ij}\sigma_{j}^{\prime T}\hat{W}_{aj}^{T}\frac{\omega_{i}^{T}}{\rho_{i}}\hat{W}_{ci}^{T}\\
+\frac{1}{4}\sum_{k=1}^{M_{xi}}\sum_{j=1}^{N}\frac{\eta_{c2i}}{M_{xi}}\sigma_{j}^{\prime ik}G_{ij}^{ik}\left(\sigma_{j}^{\prime ik}\right)^{T}\hat{W}_{aj}^{T}\frac{\left(\omega_{i}^{k}\right)^{T}}{\rho_{i}^{k}}\hat{W}_{ci}^{T},\label{eq:ActorUpdate}
\end{multline}
where $\eta_{a1i},\eta_{a2i}\in\mathbb{R}$ are positive constant
adaptation gains and $G_{\sigma i}\triangleq\sigma_{i}^{\prime}g_{i}R_{ii}^{-1}g_{i}^{T}\sigma_{i}^{\prime T}\in\mathbb{R}^{p_{Wi}\times p_{Wi}}$.
The forgetting factors $\beta_{i}$ along with the saturation in the
update laws for the least squares gain matrices in (\ref{eq:CriticUpdate})
ensure (cf. \cite{Ioannou1996}) that the least squares gain matrices
$\Gamma_{i}$ and their inverses are positive definite and bounded
for all $i\in\left\{ 1,..,N\right\} $ as 
\begin{equation}
\underline{\Gamma}_{i}\leq\left\Vert \Gamma_{i}\left(t\right)\right\Vert \leq\overline{\Gamma}_{i},\forall t\in\mathbb{R}_{\geq0},\label{eq:GammaBound}
\end{equation}
 where $\underline{\Gamma}_{i}\in\mathbb{R}$ are positive constants,
and the normalized regressors are bounded as $\left\Vert \frac{\omega_{i}}{\rho_{i}}\right\Vert \leq\frac{1}{2\sqrt{\nu_{i}\underline{\Gamma}_{i}}}.$

\section{Stability analysis}

Subtracting (\ref{eq:HJBCL*}) from (\ref{eq:deltaiHat}), the approximate
BEs can be expressed in an unmeasurable form as %
\begin{align}
\hat{\delta}_{i} & =-\omega_{i}^{T}\tilde{W}_{ci}+\frac{1}{4}\sum_{j=1}^{N}\tilde{W}_{aj}^{T}\sigma_{j}^{\prime}G_{ij}\sigma_{j}^{\prime T}\tilde{W}_{aj}\nonumber \\
 & -\frac{1}{2}\sum_{j=1}^{N}\left(W_{i}^{T}\sigma_{i}^{\prime}G_{j}-W_{j}^{T}\sigma_{j}^{\prime}G_{ij}\right)\sigma_{j}^{\prime T}\tilde{W}_{aj}-\epsilon_{i}^{\prime}f+\Delta_{i},\label{eq:deltaiunm}
\end{align}
where $\Delta_{i}\triangleq\frac{1}{2}\sum_{j=1}^{N}\left(W_{i}^{T}\sigma_{i}^{\prime}G_{j}-W_{j}^{T}\sigma_{j}^{\prime}G_{ij}\right)\epsilon_{j}^{\prime T}+\frac{1}{2}\sum_{j=1}^{N}W_{j}^{T}\sigma_{j}^{\prime}G_{j}\epsilon_{i}^{\prime T}+\frac{1}{2}\sum_{j=1}^{N}\epsilon_{i}^{\prime}G_{j}\epsilon_{j}^{\prime T}-\sum_{j=1}^{N}\frac{1}{4}\epsilon_{j}^{\prime}G_{ij}\epsilon_{j}^{\prime T}.$
Similarly, the approximate BEs evaluated at the selected points can
be expressed in an unmeasurable form as
\begin{align}
\hat{\delta}_{i}^{k} & =-\omega_{i}^{kT}\tilde{W}_{ci}+\frac{1}{4}\sum_{j=1}^{N}\tilde{W}_{aj}^{T}\sigma_{j}^{\prime ik}G_{ij}^{ik}\left(\sigma_{j}^{\prime ik}\right)^{T}\tilde{W}_{aj}+\Delta_{i}^{k}\nonumber \\
 & -\frac{1}{2}\sum_{j=1}^{N}\left(W_{i}^{T}\sigma_{i}^{\prime ik}G_{j}^{ik}-W_{j}^{T}\sigma_{j}^{\prime ik}G_{ij}^{ik}\right)\left(\sigma_{j}^{\prime ik}\right)^{T}\tilde{W}_{aj},\label{eq:deltaikunm}
\end{align}
where the constants $\Delta_{i}^{k}\in\mathbb{R}$ are defined as
$\Delta_{i}^{k}\triangleq-\epsilon_{i}^{\prime ik}f^{ik}+\Delta_{i}^{ik}$.
To facilitate the stability analysis, a candidate Lyapunov function
is defined as 
\begin{align}
V_{L} & =\sum_{i=1}^{N}V_{i}^{*}+\frac{1}{2}\sum_{i=1}^{N}\tilde{W}_{ci}^{T}\Gamma_{i}^{-1}\tilde{W}_{ci}+\frac{1}{2}\sum_{i=1}^{N}\tilde{W}_{ai}^{T}\tilde{W}_{ai}\label{eq:VL-1}
\end{align}
Since $V_{i}^{*}$ are positive definite, the bound in (\ref{eq:GammaBound})
and Lemma 4.3 in \cite{Khalil2002} can be used to bound the candidate
Lyapunov function as
\begin{equation}
\underline{v}\left(\left\Vert Z\right\Vert \right)\leq V_{L}\left(Z,t\right)\leq\overline{v}\left(\left\Vert Z\right\Vert \right),\label{eq:VLBound}
\end{equation}
where $Z=\left[x^{T},\tilde{W}_{c1}^{T},..,\tilde{W}_{cN}^{T},\tilde{W}_{a1}^{T},..,\tilde{W}_{aN}^{T}\right]^{T}\in\mathbb{R}^{2n+2N\sum_{i}p_{W_{i}}}$
and $\underline{v},\overline{v}:\mathbb{R}_{\geq0}\to\mathbb{R}_{\geq0}$
are class $\mathcal{K}$ functions. For any compact set $\mathcal{Z}\subset\mathbb{R}^{2n+2N\sum_{i}p_{W_{i}}},$
define 
\begin{gather}
\mbox{\ensuremath{\iota}}_{1}\triangleq\max_{i,j}\left(\sup_{Z\in\mathcal{Z}}\left\Vert \frac{1}{2}W_{i}^{T}\sigma_{i}^{\prime}G_{j}\sigma_{j}^{\prime T}+\frac{1}{2}\epsilon_{i}^{\prime}G_{j}\sigma_{j}^{\prime T}\right\Vert \right)\nonumber \\
\mbox{\ensuremath{\iota}}_{2}\triangleq\max_{i,j}\Bigl(\sup_{Z\in\mathcal{Z}}\Bigl\Vert\frac{\eta_{c1i}\omega_{i}}{4\rho_{i}}\left(3W_{j}\sigma_{j}^{\prime}G_{ij}-2W_{i}^{T}\sigma_{i}^{\prime}G_{j}\right)\sigma_{j}^{\prime T}\nonumber \\
\quad+\sum_{k=1}^{M_{xi}}\frac{\eta_{c2i}\omega_{i}^{k}}{4M_{xi}\rho_{i}^{k}}\left(3W_{j}^{T}\sigma_{j}^{\prime ik}G_{ij}^{ik}-2W_{i}^{T}\sigma_{i}^{\prime ik}G_{j}^{ik}\right)\left(\sigma_{j}^{\prime ik}\right)^{T}\Bigr\Vert\Bigr)\nonumber \\
\iota_{3}\triangleq\max_{i,j}\Bigl(\sup_{Z\in\mathcal{Z}}\Bigl\Vert\frac{1}{2}\sum_{i,j=1}^{N}\left(W_{i}^{T}\sigma_{i}^{\prime}+\epsilon_{i}^{\prime}\right)G_{j}\epsilon_{j}^{\prime T}\nonumber \\
\quad-\frac{1}{4}\sum_{i,j=1}^{N}\left(2W_{j}^{T}\sigma_{j}^{\prime}+\epsilon_{j}^{\prime}\right)G_{ij}\epsilon_{j}^{\prime T}\Big\Vert\Big)\nonumber \\
\iota_{4}\triangleq\max_{i,j}\left(\sup_{Z\in\mathcal{Z}}\left\Vert \sigma_{j}^{\prime}G_{ij}\sigma_{j}^{\prime T}\right\Vert \right),\:\iota_{5i}\triangleq\frac{\eta_{c1i}L_{f}\overline{\epsilon}_{i}^{\prime}}{4\sqrt{\nu_{i}\underline{\Gamma}_{i}}}\nonumber \\
\iota_{8}\triangleq\sum_{i=1}^{N}\frac{\left(\eta_{c1i}+\eta_{c2i}\right)\overline{W}_{i}\iota_{4}}{8\sqrt{\nu_{i}\underline{\Gamma}_{i}}},\:\iota_{9i}\triangleq\left(\mbox{\ensuremath{\iota}}_{1}N+\left(\eta_{a2i}+\iota_{8}\right)\overline{W}_{i}\right)\nonumber \\
\iota_{10i}\triangleq\frac{\eta_{c1i}\sup_{Z\in\mathcal{Z}}\left\Vert \Delta_{i}\right\Vert +\eta_{c2i}\max_{k}\left\Vert \Delta_{i}^{k}\right\Vert }{2\sqrt{\nu_{i}\underline{\Gamma}_{i}}}\nonumber \\
v_{l}\triangleq\frac{1}{2}\min\left(\frac{\underline{q_{i}}}{2},\frac{\eta_{c2i}\underline{c}_{xi}}{4},\frac{2\eta_{a1i}+\eta_{a2i}}{8}\right)\nonumber \\
\iota\triangleq\sum_{i=1}^{N}\left(\frac{2\iota_{9i}^{2}}{2\eta_{a1i}+\eta_{a2i}}+\frac{\iota_{10i}^{2}}{\eta_{c2i}\underline{c}_{xi}}\right)+\iota_{3},\nonumber \\
\overline{Z}\triangleq\underline{v}^{-1}\left(\overline{v}\left(\max\left(\left\Vert Z\left(t_{0}\right)\right\Vert ,\sqrt{\frac{\iota}{v_{l}}}\right)\right)\right)\label{eq:bounds}
\end{gather}
where $\underline{q_{i}}$ denote the minimum eigenvalues of $Q_{i}$
and the suprema exist since $\frac{\omega_{i}}{\rho_{i}}$ are uniformly
bounded for all $Z$, and the functions $G_{i}$, $G_{ij}$, $\sigma_{i}^{\prime}$,
and $\epsilon_{i}^{\prime}$ are continuous. In (\ref{eq:bounds}),
$L_{f}\in\mathbb{R}_{\geq0}$ denotes the Lipschitz constant such
that $\left\Vert f\left(\varpi\right)\right\Vert \leq L_{f}\left\Vert \varpi\right\Vert $
for all $\varpi\in\mathcal{Z}\cap\mathbb{R}^{n}.$ The sufficient
conditions for UUB convergence are derived based on the subsequent
stability analysis as
\begin{gather}
\underline{q_{i}}>2\iota_{5i},\quad\eta_{c2i}\underline{c}_{xi}>2\iota_{5i}+\iota_{2}\zeta N+\eta_{a1i},\nonumber \\
2\eta_{a1i}+\eta_{a2i}>4\iota_{8}+\frac{2\iota_{2}N}{\zeta},\label{eq:GainCond}
\end{gather}
where $\zeta\in\mathbb{R}$ is a known positive adjustable constant.

Since the NN function approximation error and the Lipschitz constant
$L_{f}$ depend on the compact set that contains the state trajectories,
the compact set needs to be established before the gains can be selected
using (\ref{eq:GainCond}). Based on the subsequent stability analysis,
an algorithm is developed to compute the required compact set (denoted
by $\mathcal{Z}$) based on the initial conditions. In Algorithm \ref{alg:Gain-Selection},
the notation $\left\{ \varpi\right\} _{i}$ for any parameter $\varpi$
denotes the value of $\varpi$ computed in the $i^{th}$ iteration.
\begin{algorithm}
\uline{First iteration:}

Given an upper bound $z\in\mathbb{R}_{\geq0}$ on $Z\left(t_{0}\right)$
such that $\left\Vert Z\left(t_{0}\right)\right\Vert <z$, let $\mathcal{Z}_{1}\triangleq\left\{ \xi\in\mathbb{R}^{2n+2N\sum_{i}\left\{ p_{W_{i}}\right\} _{1}}\mid\left\Vert \xi\right\Vert \leq\underline{v}^{-1}\left(\overline{v}\left(z\right)\right)\right\} $.
Using $\mathcal{Z}_{1},$ compute the bounds in (\ref{eq:bounds})
and select the gains according to (\ref{eq:GainCond}). If $\left\{ \sqrt{\frac{\iota}{v_{l}}}\right\} _{1}\leq z$,
set $\mathcal{Z}=\mathcal{Z}_{1}$ and terminate.

\uline{Second iteration:}

If $z<\left\{ \sqrt{\frac{\iota}{v_{l}}}\right\} _{1}$, let $\mathcal{Z}_{2}\triangleq\left\{ \xi\in\mathbb{R}^{2n+2N\sum_{i}\left\{ p_{W_{i}}\right\} _{1}}\mid\left\Vert \xi\right\Vert \leq\underline{v}^{-1}\left(\overline{v}\left(\left\{ \sqrt{\frac{\iota}{v_{l}}}\right\} _{1}\right)\right)\right\} $.
Using $\mathcal{Z}_{2},$ compute the bounds in (\ref{eq:bounds})
and select the gains according to (\ref{eq:GainCond}). If $\left\{ \sqrt{\frac{\iota}{v_{l}}}\right\} _{2}\leq\left\{ \sqrt{\frac{\iota}{v_{l}}}\right\} _{1}$,
set $\mathcal{Z}=\mathcal{Z}_{2}$ and terminate.

\uline{Third iteration:}

If $\left\{ \sqrt{\frac{\iota}{v_{l}}}\right\} _{2}>\left\{ \sqrt{\frac{\iota}{v_{l}}}\right\} _{1}$,
increase the number of NN neurons to $\left\{ p_{Wi}\right\} _{3}$
to ensure $\left\{ L_{f}\right\} _{2}\left\{ \overline{\epsilon}_{i}^{\prime}\right\} _{3}\leq\left\{ L_{f}\right\} _{2}\left\{ \overline{\epsilon}_{i}^{\prime}\right\} _{2},\forall i=1,..,N.$
These adjustments ensure $\left\{ \iota\right\} _{3}\leq\left\{ \iota\right\} _{2}$.
Set $\mathcal{Z}=\left\{ \xi\in\mathbb{R}^{2n+2N\sum_{i}\left\{ p_{W_{i}}\right\} _{3}}\mid\left\Vert \xi\right\Vert \leq\underline{v}^{-1}\left(\overline{v}\left(\left\{ \sqrt{\frac{\iota}{v_{l}}}\right\} _{2}\right)\right)\right\} $
and terminate.

\caption{\label{alg:Gain-Selection}Gain Selection}
\end{algorithm}
{} Since the constants $\iota$ and $v_{l}$ depend on $L_{f}$ only
through the product $L_{f}\overline{\epsilon}_{i}^{\prime}$, Algorithm
\ref{alg:Gain-Selection} ensures that
\begin{equation}
\sqrt{\frac{\iota}{v_{l}}}\leq\frac{1}{2}\mbox{diam}\left(\mbox{\ensuremath{\mathcal{Z}}}\right),\label{eq:iotaCond}
\end{equation}
where $\mbox{diam\ensuremath{\left(\mathcal{Z}\right)}}$ denotes
the diameter of the set $\mathcal{Z}$.
\begin{thm}
Provided Assumption \ref{ass:CLW} holds and the control gains satisfy
the sufficient conditions in (\ref{eq:GainCond}), where the constants
in (\ref{eq:bounds}) are computed based on the compact set $\mathcal{Z}$
selected using Algorithm \ref{alg:Gain-Selection}, the controllers
in (\ref{eq:Vu}) along with the adaptive update laws in (\ref{eq:CriticUpdate})
and (\ref{eq:ActorUpdate}) ensure that the state $x$, the value
function weight estimation errors $\tilde{W}_{ci}$ and the policy
weight estimation errors $\tilde{W}_{ai}$ are UUB, resulting in UUB
convergence of the policies $\hat{u}_{i}$ to the feedback-Nash equilibrium
policies $u_{i}^{*}$.\end{thm}
\begin{IEEEproof}
The derivative of the candidate Lyapunov function in (\ref{eq:VL-1})
along the trajectories of (\ref{eq:Dynamics}), (\ref{eq:CriticUpdate}),
and (\ref{eq:ActorUpdate}) is given by%
\begin{align}
\dot{V}_{L} & =\sum_{i=1}^{N}\left(\nabla_{x}V_{i}^{*}\left(f+\sum_{j=1}^{N}g_{j}u_{j}\right)\right)\nonumber \\
 & +\sum_{i=1}^{N}\tilde{W}_{ci}^{T}\left(\frac{\eta_{c1i}\omega_{i}}{\rho_{i}}\hat{\delta}_{i}+\frac{\eta_{c2i}}{M_{xi}}\sum_{i=1}^{M_{xi}}\frac{\omega_{i}^{k}}{\rho_{i}^{k}}\hat{\delta}_{i}^{k}\right)\nonumber \\
 & -\frac{1}{2}\sum_{i=1}^{N}\tilde{W}_{ci}^{T}\left(\beta_{i}\Gamma_{i}^{-1}-\eta_{c1i}\frac{\omega_{i}\omega_{i}^{T}}{\rho_{i}^{2}}\right)\tilde{W}_{ci}\nonumber \\
 & -\sum_{i=1}^{N}\tilde{W}_{ai}^{T}\Bigg(-\eta_{a1i}\left(\hat{W}_{ai}^{T}-\hat{W}_{ci}^{T}\right)-\eta_{a2i}\hat{W}_{ai}^{T}\nonumber \\
 & +\frac{1}{4}\sum_{j=1}^{N}\eta_{c1i}\hat{W}_{ci}^{T}\frac{\omega_{i}}{\rho_{i}}\hat{W}_{aj}^{T}\sigma_{j}^{\prime}G_{ij}\sigma_{j}^{\prime T}\nonumber \\
 & +\frac{1}{4}\sum_{k=1}^{M_{xi}}\sum_{j=1}^{N}\frac{\eta_{c2i}}{M_{xi}}\hat{W}_{ci}^{T}\frac{\omega_{i}^{k}}{\rho_{i}^{k}}\hat{W}_{aj}^{T}\sigma_{j}^{\prime ik}G_{ij}^{ik}\left(\sigma_{j}^{\prime ik}\right)^{T}\Bigg).\label{eq:VLDot1}
\end{align}
Substituting the unmeasurable forms of the BEs from (\ref{eq:deltaiunm})
and (\ref{eq:deltaikunm}) into (\ref{eq:VLDot1}) and using the triangle
inequality, the Cauchy-Schwarz inequality and Young's inequality,
the Lyapunov derivative in (\ref{eq:VLDot1}) can be bounded as
\begin{align}
 & \dot{V}\leq-\sum_{i=1}^{N}\frac{q_{i}}{2}\left\Vert x\right\Vert ^{2}-\sum_{i=1}^{N}\frac{\eta_{c2i}\underline{c}_{xi}}{2}\left\Vert \tilde{W}_{ci}\right\Vert ^{2}\nonumber \\
 & -\sum_{i=1}^{N}\left(\frac{2\eta_{a1i}+\eta_{a2i}}{4}\right)\left\Vert \tilde{W}_{ai}\right\Vert ^{2}+\sum_{i=1}^{N}\iota_{9i}\left\Vert \tilde{W}_{ai}\right\Vert \nonumber \\
 & +\sum_{i=1}^{N}\iota_{10i}\left\Vert \tilde{W}_{ci}\right\Vert -\sum_{i=1}^{N}\left(\frac{q_{i}}{2}-\iota_{5i}\right)\left\Vert x\right\Vert ^{2}\nonumber \\
 & -\sum_{i=1}^{N}\left(\frac{\eta_{c2i}\underline{c}_{xi}}{2}-\left(\iota_{5i}+\frac{\iota_{2}\zeta_{2}N}{2}+\frac{\eta_{a1i}}{2}\right)\right)\left\Vert \tilde{W}_{ci}\right\Vert ^{2}\nonumber \\
 & +\sum_{i=1}^{N}\left(\frac{2\eta_{a1i}+\eta_{a2i}}{4}-\iota_{8}-\frac{\iota_{2}N}{2\zeta_{2}}\right)\left\Vert \tilde{W}_{ai}\right\Vert ^{2}\nonumber \\
 & +\iota_{3}.\label{eq:VLDot2}
\end{align}
Provided the sufficient conditions in (\ref{eq:GainCond}) hold, completing
the squares in (\ref{eq:VLDot2}), the bound on the Lyapunov derivative
can be expressed as
\begin{align}
\dot{V} & \leq-\sum_{i=1}^{N}\frac{\underline{q_{i}}}{2}\left\Vert x\right\Vert ^{2}-\sum_{i=1}^{N}\frac{\eta_{c2i}\underline{c}_{xi}}{4}\left\Vert \tilde{W}_{ci}\right\Vert ^{2}\nonumber \\
 & -\sum_{i=1}^{N}\left(\frac{2\eta_{a1i}+\eta_{a2i}}{8}\right)\left\Vert \tilde{W}_{ai}\right\Vert ^{2}+\iota,\nonumber \\
 & <-v_{l}\left\Vert Z\right\Vert ^{2},\quad\forall\left\Vert Z\right\Vert >\sqrt{\frac{\iota}{v_{l}}}.\label{eq:VLDot3}
\end{align}
Using (\ref{eq:VLBound}), (\ref{eq:iotaCond}), and (\ref{eq:VLDot3}),
Theorem 4.18 in \cite{Khalil2002} can be invoked to conclude that
$\lim\sup_{t\to\infty}\left\Vert Z\left(t\right)\right\Vert \leq\underline{v}^{-1}\left(\overline{v}\left(\sqrt{\frac{\iota}{v_{l}}}\right)\right).$
Furthermore, the system trajectories are bounded as $\left\Vert Z\left(t\right)\right\Vert \leq\overline{Z}$
for all $t\in\mathbb{R}_{\geq0}$. 

The error between the feedback-Nash equilibrium policies and the approximate
policies can be expressed as 
\[
\left\Vert u_{i}^{*}-\hat{u}_{i}\right\Vert \leq\frac{1}{2}\left\Vert R_{ii}\right\Vert \overline{g_{i}}\overline{\sigma}_{i}^{\prime}\left(\left\Vert \tilde{W}_{ai}\right\Vert +\bar{\epsilon}_{i}^{\prime}\right),
\]
for all $i=1,..,N$, where $\overline{g_{i}}\triangleq\sup_{x}\left\Vert g_{i}\left(x\right)\right\Vert $.
Since the weights $\tilde{W}_{ai}$ are UUB, UUB convergence of the
approximate policies to the feedback-Nash equilibrium policies is
obtained.\end{IEEEproof}
\begin{rem}
The closed-loop system analyzed using the candidate Lyapunov function
in (\ref{eq:VL-1}) is a switched system. The switching happens when
the least squares regression matrices $\Gamma_{i}$ reach their saturation
bound. Similar to least squares-based adaptive control (cf. \cite{Ioannou1996}),
(\ref{eq:VL-1}) can be shown to be a common Lyapunov function for
the regression matrix saturation. Since (\ref{eq:VL-1}) is a common
Lyapunov function, (\ref{eq:VLBound}), (\ref{eq:iotaCond}), and
(\ref{eq:VLDot3}) establish UUB convergence of the switched system.
\end{rem}

\section{Conclusion}

A concurrent learning-based adaptive approach is developed to determine
the feedback-Nash equilibrium solution to an $N$-player nonzero-sum
game online. The solutions to the associated coupled HJ equations
and the corresponding feedback-Nash equilibrium policies are approximated
using parametric universal function approximators. Based on the system
dynamics, the Bellman errors are evaluated at a set of preselected
points in the state-space. The value function and the policy weights
are updated using a concurrent learning-based least squares approach
to minimize the instantaneous BEs and the BEs evaluated at the preselected
points. 

Unlike traditional approaches that require a PE condition for convergence,
UUB convergence of the value function and policy weights to their
true values, and hence, UUB convergence of the policies to the feedback-Nash
equilibrium policies, is established under weaker rank conditions
using a Lyapunov-based analysis. The developed result relies on a
sufficient condition on the minimum eigenvalue of a time-varying regressor
matrix. While this condition can be heuristically satisfied by choosing
enough points, and can be easily verified online, it can not, in general,
be guaranteed a priori. Furthermore, finding a sufficiently good basis
for value function approximation is, in general, nontrivial and can
be achieved only through prior knowledge or trial and error. Future
research will focus on extending the applicability of the developed
technique by investigating the aforementioned challenges.\bibliographystyle{IEEEtran}
\bibliography{ncr,master}

\end{document}